# Direct Measurement of the Exciton Binding Energy and Effective Masses for Charge carriers in an Organic-Inorganic Tri-halide Perovskite

Atsuhiko Miyata[†1], Anatolie Mitioglu[†1], Paulina Plochocka[1], Oliver Portugall[1], Jacob Tse-Wei Wang[2], Samuel D. Stranks[2], Henry J. Snaith[2] and Robin J. Nicholas[2]*

[1] Laboratoire National des Champs Magnetiques Intenses, CNRS-UJF-UPS-INSA, 143 Avenue de Rangueil, 31400 Toulouse, France

[2] University of Oxford, Clarendon Laboratory, Parks Road, Oxford, OX1 3PU, United Kingdom

* Correspondence to: r.nicholas@physics.ox.ac.uk
[†] *the authors equally contributed*

**Solar cells based on the organic-inorganic tri-halide perovskite family of materials have shown remarkable progress recently, offering the prospect of low-cost solar energy from devices that are very simple to process. Fundamental to understanding the operation of these devices is the exciton binding energy, which has proved both difficult to measure directly and controversial. We demonstrate that by using very high magnetic fields it is possible to make an accurate and direct spectroscopic measurement of the exciton binding energy, which we find to be only 16 meV at low temperatures, over three times smaller than has been previously assumed. In the room temperature phase we show that the binding energy falls to even smaller values of only a few millielectronvolts, which explains**

**their excellent device performance due to spontaneous free carrier generation following light absorption. Additionally, we determine the excitonic reduced effective mass to be 0.104$m_e$ (where $m_e$ is the electron mass), significantly smaller than previously estimated experimentally but in good agreement with recent calculations. Our work provides crucial information about the photophysics of these materials, which will in turn allow improved optoelectronic device operation and better understanding of their electronic properties**

## Introduction

The recent rapid development of perovskite solar cells is revolutionizing the photovoltatic (PV) research field, with the latest certified power conversion efficiencies reaching over 20% [1]. Initially developed from the concept of the nanostructured excitonic solar cell where there is no requirement for long range charge or exciton diffusion [1-8], it has now become clear that due to the remarkable properties of the inorganic-organic perovskite family of materials $ABX_3$ (A=$CH_3NH_3^+$; B=$Pb^{2+}$; and X = $Cl^-$, $I^-$ and/or $Br^-$) these cells [2-10] are capable of operating in a comparable configuration and with comparable performance to the best inorganic semiconductors, [7, 9,10] where the solid absorber layer is sandwiched between n- and p-type charge selective contacts in a planar heterojunction configuration [7-8]. Despite this success several fundamental properties of the organic lead tri-halide perovskites remain controversial and poorly known. In particular the binding energy of the excitons (R*), bound electron-hole pairs that are the primary photoexcited species created in the absorption process, is vital to understanding the way that the cells function. The operating mechanisms depend upon what fraction of excitons dissociate in the bulk material, giving rise to free charge transport, or what fraction need to be dissociated at heterojunctions within the cells. Knowledge of the true exciton binding energy is also crucial for interpreting spectroscopic measurements based on these materials, such as time-resolved spectroscopy. Values for R* reported in the literature cover a broad range from 2 to 55 meV[11-17], with the larger values being initially adopted and a growing number of reports suggesting a wide range of lower values[15-17]. In addition basic parameters such as the effective masses of electrons and holes also remain to be directly measured in the archetypical material $CH_3NH_3PbI_3$. A number of calculations of the band structure in the literature are able to reproduce the observed band gaps[14,18,19] and these suggest that the conduction and valence bands are essentially isotropic and symmetrical[14,18,19]. A direct measurement of the exciton binding energy and effective masses is therefore crucial for our current understanding and for future development of this remarkable class of materials.

In this work we describe the use of very high field inter-band magneto-absorption studies which allow us, unlike previous PhotoLuminescence measurements[20], to make an accurate study of the family of free exciton states, which are the relevant excitations created in PV devices. In the low temperature orthorhombic phase we establish that the exciton binding energy is only 16 meV, which is significantly smaller than has been previously assumed. We also investigate the room temperature tetragonal phase, which occurs above 160K[11,19], where we find the striking result that the binding energy falls to only a few millielectron volts. Our measurements give an accurate and independent value for the reduced effective mass and in addition, show that the essentially symmetric and isotropic conduction and valence bands of the organic-inorganic lead tri-halide perovskites makes them model semiconductors to demonstrate the optical properties of excitons in a high magnetic field.

**Magnetic field dependent optical absorption.** In Fig. 1 we summarize the results of transmission measurements of a ~300-nm thick polycrystalline film of $CH_3NH_3PbI_3$, deposited directly on a glass substrate, measured at 2 K in magnetic fields of up to 65 T in the Faraday configuration using a long (500ms) pulsed field magnet. Close to the band edge (~1.6 eV), the spectra are dominated by the hydrogen-like exciton states and at higher energies the free carrier behaviour of the conduction and valence bands gives rise to a series of interband transitions between the van Hove singularities at the bottom of the Landau levels with energies given by:

$$E(B) = E_g + (N + 1/2)\hbar\omega_c \pm 1/2\, g_{eff}\mu_B B,$$

where $E_g$ is the energy gap, N=0,1,2,3.. is the Landau quantum number, $\omega_c = eB/m^*$, B is the applied magnetic field, e is the elementary charge, $m^*$ is the reduced effective mass of the exciton given by, $1/m^* = 1/m^*_e + 1/m^*_h$, where $m_e$ and $m_h$ are the electron and hole effective masses respectively, $g_{eff}$ is the effective g-factor for the Zeeman splitting, and $\mu_B$ is the Bohr magneton. In the present case we use unpolarised light so that the last term is ignored. In contrast to the early magneto-optical studies on this material[12,13] our spectra show a clearly resolved 1s exciton at 1.64 eV, which has a large diamagnetic shift, significantly larger than the peak width and larger than the values observed previously[12,13]. In addition, the spectra we show in Fig. 1 demonstrate several important new features as the field is increased. By 65T there is a clear sequence of 5 well-resolved Landau level transitions which develop in magnetic fields above 30T. Furthermore, a small shoulder develops on the high energy side of the 1s exciton peak which we identify as the 2s exciton absorption. The appearance of the Landau levels and the 2s exciton state can be seen more clearly by taking the ratio of the high field spectra to zero field as shown in Fig. 1b and 1c, where the resonant absorptions become minima in the ratio of transmission. This shows that the 2s state is clearly visible as a weak but gradually growing absorption in the range 10 - 35 T. Above 20 T a second absorption grows much more rapidly which we attribute to a combination of the 2p exciton and the first of the free-electron inter-band Landau level transitions. The 2p exciton transition is forbidden at zero magnetic field but becomes allowed at high fields due to the re-construction of the hydrogenic energy levels which occurs once the cyclotron energy exceeds the exciton binding energy ($\hbar\omega_c > R^*$) [21]. At high fields the conventional atomic quantum numbers (n, l) are no longer valid and this state is renamed as (1,0) corresponding to the strongest bound state associated with the N=1 Landau level.

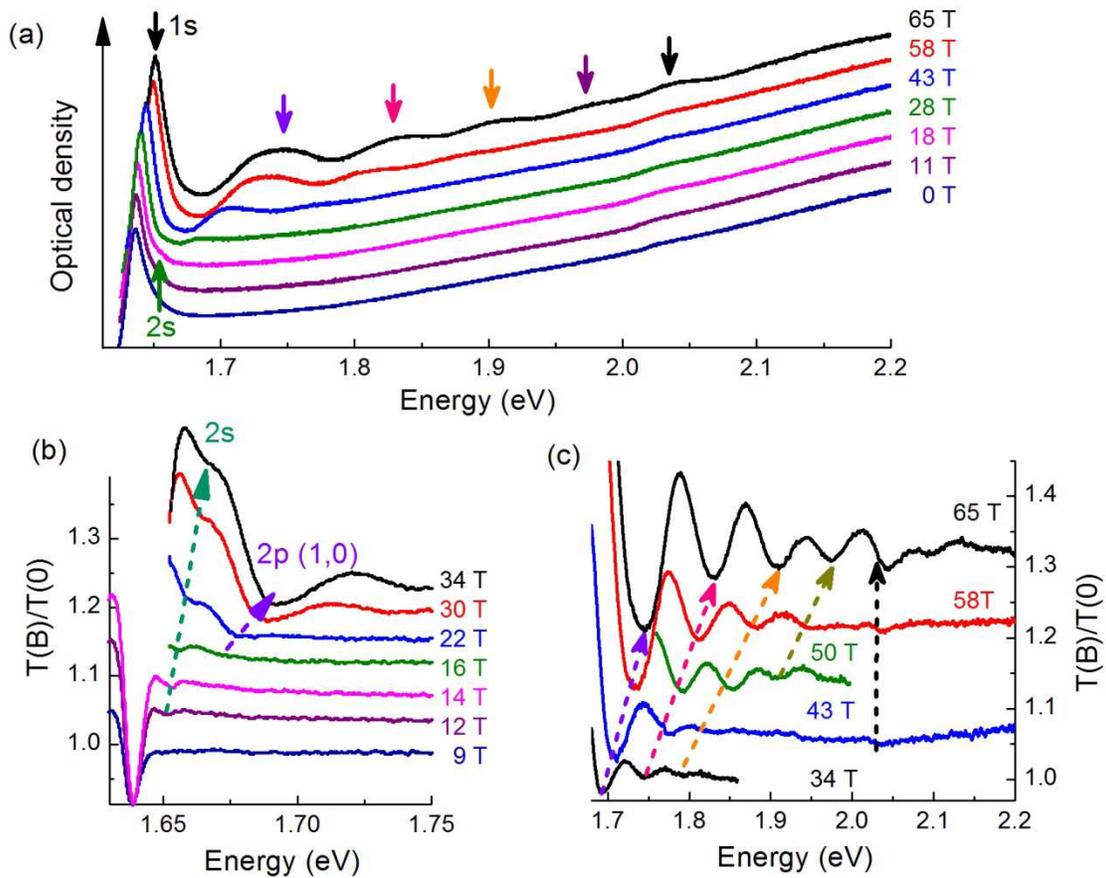

**Figure 1| Magnetic field dependence of the optical density for the perovskite** $CH_3NH_3PbI_3$

(a) A sequence of optical density (Log(1/transmission)) spectra measured during a single pulse

[5]

of the magnetic field. To improve the resolution of smaller field dependent features (b) and (c) show sequences of ratios of the transmission in magnetic field T(B) to that measured at zero field T(0), where the resonant absorption features correspond to minima. For easier comparison spectra are offset. The feature highlighted at 2.03 eV is a band edge absorption from a previously undetected higher energy band edge and will be the subject of a future publication.

The separation between the sequence of Landau levels visible directly in the absorption spectrum at 65T immediately allows us to estimate the reduced effective mass to be m* ≈ 0.1 $m_e$ to within a few per cent accuracy, while the separation of the 1s and 2s states at low field (15 meV at 10 T) allows us to estimate an excitonic binding energy, R*, of order 20 meV, which is consistent with the condition $\hbar\omega_c$ > R* for fields above 14T in agreement with our preliminary analysis of fig. 1b. We will present a more detailed fitting to the full family of transitions below, which allows us to deduce a precise value of R* being 16±2 meV. The values for the effective mass are in good agreement with recent calculations[14,18] and somewhat smaller than previous experimental estimates. The excitonic binding energy is much smaller than the early spectroscopic values of 30-50 meV although in good agreement with a recently calculated theoretical value which takes into account the frequency dependent refractive index of the perovskite.

**Magneto-optical analysis.** We now perform a full fitting, where we also include data taken up to 150T using a fast pulse single turn magnetic field system at a fixed photon energy. In order to extract accurate values for the effective mass and exciton binding energy from the magnetic field dependence of the transition energies it is important to use full numerical calculations for the magnetic field dependent transitions as the data cover a wide range from the low to the high magnetic field limits and no analytical solution exists for the hydrogen atom in a high magnetic field. To do this we use the values calculated by Makado and Magill[21] which scale the most strongly bound excitonic energy levels $E_{n,0}(\gamma)$ by the use of the dimensionless parameter $\gamma=\hbar\omega_c/2R^*$. This allows us to fit the complete range of magnetic field values. In addition to the excitonic transitions, the higher energy transitions are known[22] to become dominated by simple interband transitions between free carrier Landau levels as described above. Fig. 2 shows the full fan diagram of measured transition energies, together with the calculated transition energies for the excitonic and free electron transitions and a schematic of the optically allowed transitions observed. The values for the effective mass and exciton binding energy are then adjusted to globally fit the data. In practice the two parameters are dominated by very different parts of the data set and are not strongly interdependent.

[6]

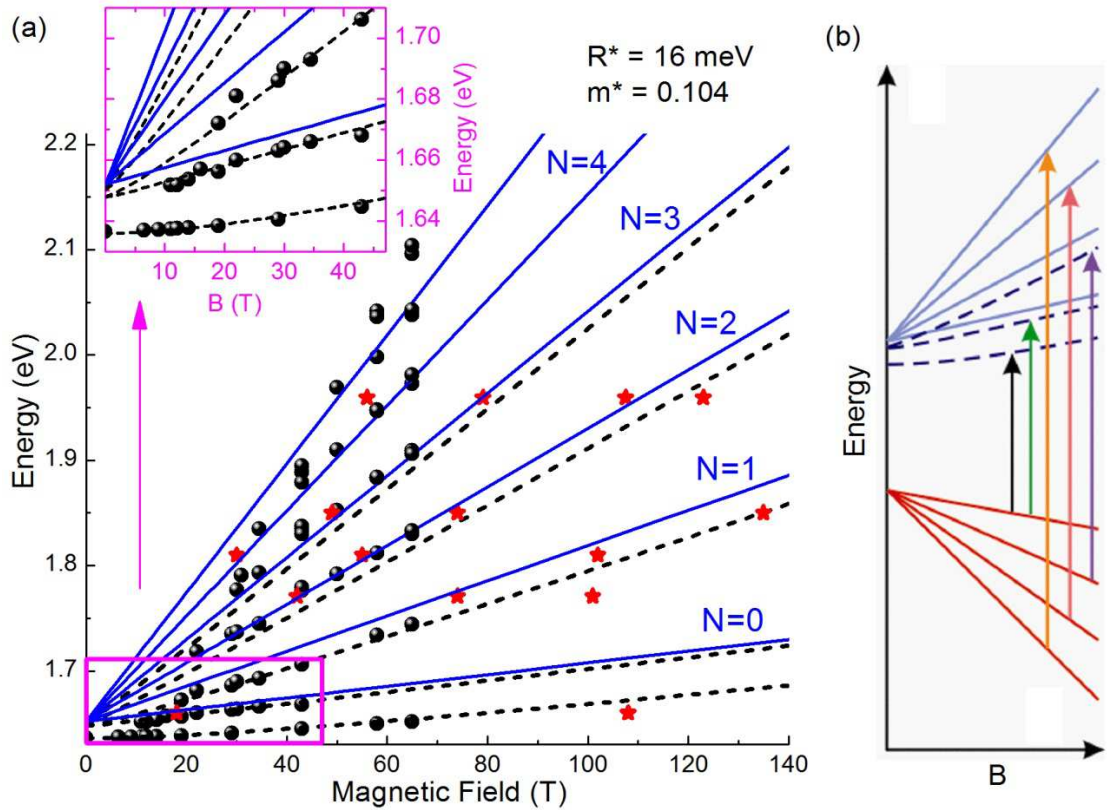

**Figure 2 Energy 'Fan' diagrams.** (**a**) Full fan using data from long pulse fixed field spectra (black circles) and fixed energy fast field sweep data (red stars). The calculated transition energies are shown for the free electron and hole levels (solid lines) and the excitonic transitions (dashed lines). Inset to (a) lower fields measured using fixed field spectra. (b) shows a schematic of the energy levels and transitions between the free electron and hole levels (solid lines) and the excitonic transitions (dashed lines).

The observation of the 2s transition places strong constraints upon the exciton binding energy, while the slope and separation of the high field, high quantum number Landau levels strongly constrain the reduced effective mass value. Further confirmation of the exciton binding energy comes from our simultaneous fitting of the diamagnetic shift of the 1s state, as shown in the inset to Fig. 2a, where the binding energy is now also strongly constrained by the accurately determined effective mass. We conclude that the excitonic transitions dominate for the N=0 landau level (1s, 2s) and the N=1 (1,0) level up to ~50T, and free electrons dominate for N=1 above 50T and for all higher Landau levels. In the intermediate region a weak splitting of the N=1 level can just be detected where both transitions are occurring. This clear demonstration of the interchange between excitonic and interband free carrier transitions is usually masked in more conventional semiconductors by complications brought about by additional degeneracies, such as the light and heavy holes present in III-V or II-VI materials[22,23] and is an illustration that in addition to their potential in applications, the organic-inorganic metal tri-halide perovskites can also act as excellent model semiconductors. A further interesting observation from the fan charts in Fig. 2 is that the Landau level separations are not significantly decreasing at higher energies (and in a related observation the Landau level fans show good linearity). This suggests that the approximation of a constant effective mass is good for a wide range of band energies and non-parabolicity effects are relatively small compared with some predictions[24].

[8]

The value deduced for the effective mass of m*=0.104±0.003 $m_e$ is remarkably close to that of 0.099 $m_e$ predicted recently by Menendez et al [14] who have adjusted the amount of exchange coupling in order to match the experimentally measured values of the band gap, and Umari et al[18], who find 0.11 me. We find the exciton binding energy (R*) to be 16±2 meV, in contrast to the earlier values of 37 to 50 meV reported in [12, 13], also determined at 4K, which were deduced by fitting only the 1s state without an independent measurement of the effective mass and using only a low magnetic field approximation and with much poorer experimental resolution. Our value here is also strongly supported by the extrapolation of the free electron transitions to zero magnetic field, which fixes the excitonic continuum.

Several authors [14-17] have pointed out that using lower mass values (0.1 $m_e$) and depending on whether the low ($\varepsilon = 25.7$) or high ($\varepsilon = 5.6$) frequency dielectric constant is used[18, 24] the conventional Wannier-Mott Hydrogenic model gives values for the excitonic binding energy ($R^*=m^*e^4/2\hbar^2\varepsilon^2$) anywhere from 2 -50 meV. In practice the exciton binding energies are comparable to several of the phonon modes[25] and so it is likely that the appropriate dielectric constant at the equivalent frequency will be intermediate between the low and high frequency limits as discussed by Huang and Lambrecht [23] for $CsSnX_3$ perovskites. Our values of R* and m* suggest a value of $\varepsilon \approx 9$ assuming the hydrogenic model. In fact Even et al [15] have recently argued from fitting the lineshape of the low temperature absorption that the exciton binding energy should be on the order of 15 meV, and that as a result of the new rotational motion of the $CH_3NH_3^+$ cations in the high temperature phase there is an additional contribution to the dielectric screening which makes the exciton binding energy fall discontinuously to 5 meV at the transition to the high temperature phase where the photovoltaic cells operate. In a similar analysis of the temperature dependent absorption, Yamada et al[16] conclude that the exciton binding energy decreases continuously from 30 meV at 13K to 6 meV at 300K.

[9]

**High Temperature Phase**. A significant complication in understanding the band structure of the organic-inorganic lead tri-halide perovskite is the presence of structural phase transitions from cubic (T>350K) to tetragonal (T>145K) to orthorhombic (low T) which result in changes in band structure and band gap[11,19]. The tetragonal (room temperature) and orthorhombic phases have similar band structures, with almost symmetric and isotropic direct band gaps at the Γ-point, with the band gap increasing by ~100 meV at the phase transition to the orthorhombic phase[11,15, 16, 27], although the magnitude of the band gap change is dependent on the growth process[15, 16]. Although the band structures are very similar the phase transition may be expected to produce significant changes in the phonon structure and consequently may significantly affect the dielectric constant and hence the exciton binding energy as discussed above.

In Fig. 3, we extend the interband magneto-optical spectra up to 150T using a fast pulse single turn magnetic field system, which shows the magnetic field dependent transmission for a series of temperatures at a fixed photon energy. From this data it is very easy to observe the influence of the lower temperature phase change. Two strong transmission minima can be observed in the low temperature data which begin to move towards lower field values as the temperature increases. This is because the perovskite band gap increases [13] with increasing temperature, causing the interband magneto-optical transitions at a fixed energy to shift towards lower magnetic field values. Above 140K there is a sudden rapid shift upwards of the resonance fields caused by the structural phase transition to the high temperature tetragonal phase, which has a lower band gap. We then repeat the magneto-optical study in this tetragonal phase using the fast pulse system as we show in fig. 3b. At these temperatures, significantly fewer resonances are resolved, but it is still possible to observe several different inter-Landau level transitions and it is possible to construct a good fan diagram(fig. 4b). This shows that the effective mass is essentially the same (m*=0.104±0.005 $m_e$) as at low temperature in the orthorhombic phase, although the band gap is slightly (~50 meV) smaller than at 2K.

[10]

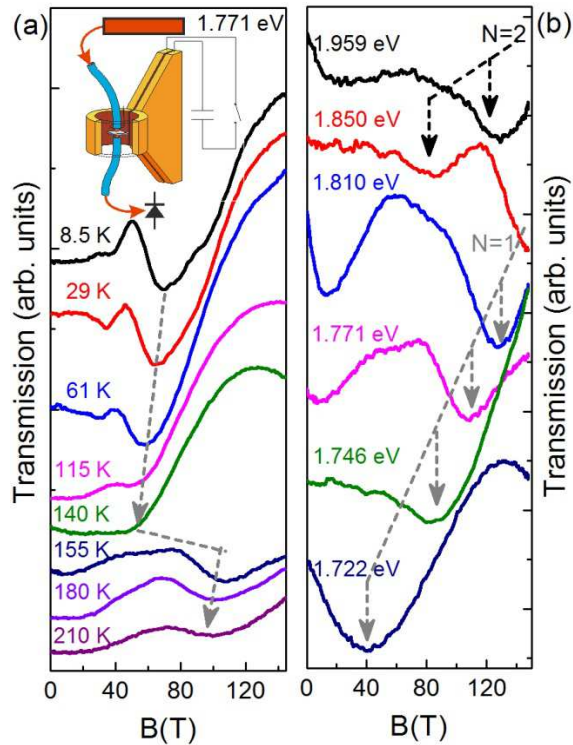

**Figure 3 Single turn coil results.** Plots of the magneto-transmission measured using a single turn fast pulse magnetic field, with schematic of the experimental system. For better comparison spectra are offset. (a) shows the temperature dependence with the arrows indicating the position of the N=1 inter Landau level transition. The inset shows a schematic of the single turn experimental system showing the coil and its firing circuit, the sample and the optical fiber

illumination and collection system. (b) shows the transmission spectra in the tetragonal (intermediate temperature) phase with spectra measured for different wavelengths in the temperature range 155 – 190 K. The linked arrows show the positions of the N=1 and N=2 transitions.

We show the optical density spectra taken in the long pulse system in Fig. 4a, which shows that the 1s exciton has considerably more broadening and as a result it is much harder to make an accurate analysis of the diamagnetic shift of the 1s transition energy, in comparison to that of the lower temperature phase. The exciton binding energy is more difficult to fit precisely because the absolute energy of the 1s exciton state is less certain and would need to be fitted with a knowledge of the scattering processes and the dielectric function. A detailed examination of the 1s exciton absorption spectra, as in fig. 4a, shows that there is an anomalous behavior in which the transition becomes much more resolved at higher magnetic fields and actually moves down in energy at first (dashed line, fig. 4a). Both the increase in intensity and the reduction in transition energy with increasing field suggest that the exciton binding energy is increasing with the magnetic field. We attribute this to a decrease in dielectric constant resulting from the additional magnetic binding of the exciton.

Using only the high field (B>50 T) spectra where Landau levels can be observed and the 1s exciton peak is well formed, we estimate a binding energy on the order of 10-12 meV *in high magnetic fields*. We extrapolate the high energy free carrier Landau levels to a band edge energy very close to the apparent 1s exciton peak at B=0 which allows us to conclude that the exciton binding energy is much smaller at zero magnetic field, with a value less than our measurement uncertainty of a few meV. This strongly supports the suggestion by Even et al[15] that the binding energy is reduced to values of order 5 meV above the phase transition to the orthorhombic phase and the recent analysis of Yamada et al [16] that there is a decrease to around 6meV at room temperature.

[12]

The overall picture is that there is a critical collapse of the exciton binding energy as a function of both increasing temperature and decreasing magnetic field. As the temperature increases any fall in binding energy decreases the frequency of motion of the electron and hole bound in the exciton leading to increased contributions to the dielectric constant from the many phonon modes and molecular rotations present in these materials. The fall in binding energy leads to a further increase in dielectric constant and the binding energy collapses to a value close to that predicted by using the low frequency dielectric constant, giving a value of a few meV. When a magnetic field is applied the process is partly reversed due to the additional cyclotron motion which increases the binding energies. Similarly low exciton binding energies due to a frequency dependent dielectric constant have been predicted [26] in the $CsSnX_3$ (X = I, Br, Cl) perovskite halide semiconductors which have very similar band and crystal structures to the organic-inorganic lead tri-halide perovskites studied here, previously leading to significant experimental controversy [26,28]. An immediate consequence of the temperature and magnetic field dependence of the exciton binding energy is that the interpretation of the earlier literature measurements as giving an exciton binding energy of 30-50 meV, which are based on the assumption of a constant binding energy as a function of magnetic field [12,13], or temperature [11], will be invalidated. Hence, our measurements show that the exciton binding energies are much smaller than previously concluded and explain why the properties of the organic-inorganic perovskites in the room temperature phase will be dominated by free carrier behavior as suggested by more recent time resolved studies[17,20,25,27,30,31].

[13]

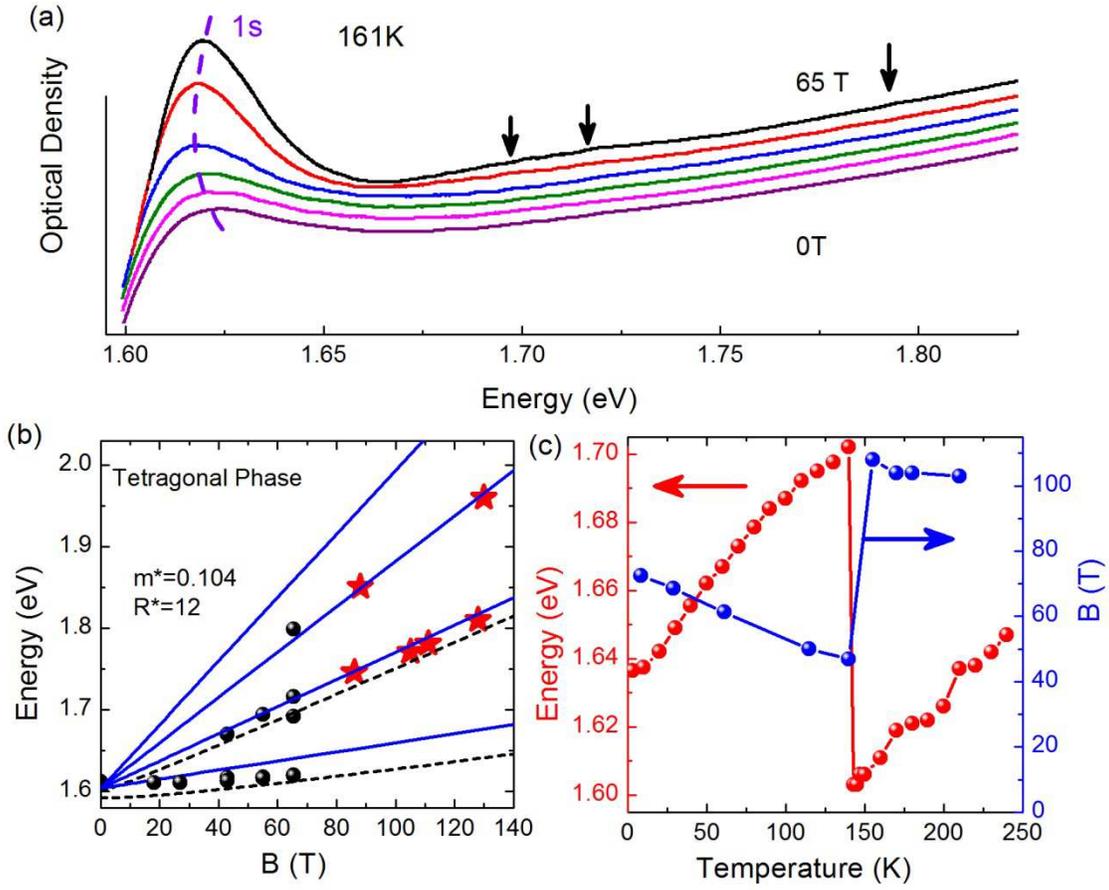

**Figure 4| Transmission in high temperature tetragonal phase** (**a**) optical density, with the dashed line showing the anomalous behavior of the 1s exciton transition. (**b**) fan diagram. (c) Temperature dependent band gaps and resonance positions for the N=1 inter-Landau level transition.

We also show in Fig. 4 the temperature dependence of the magnetic field position of the N=1 Landau level resonance, compared with the temperature dependence of the band gap as deduced from the film absorption. The sudden shift in resonance position of 63T at the phase transition temperature can be shown to be equivalent to a change in band gap of 105 meV from the slope of the N=1 resonance (1.67 meV/T), consistent with the 100 meV change in band gap deduced from absorption. This confirms that the magneto-spectroscopy is measuring the same fundamental band structure as the optical absorption.

[14]

**Conclusions**

Our basic conclusion is that the excitonic binding energy in the low temperature phase of the organic-inorganic perovskite, $CH_3NH_3PbI_3$ is much smaller (16 ± 2 meV) than has previously been estimated and is comparable to conventional III-V semiconductors with a similar band gap[29]. Our measurements in the room temperature phase suggest that there is a critical collapse of the exciton binding energies at higher temperatures, as has been predicted by Even et al [15], due to the frequency dependent dielectric constant. By room temperature the appropriate binding energy will be only a few milli-electron-volts and the photovoltaic device performance is essentially a free carrier phenomenon. This result conclusively shows that the very impressive performance of PV devices using this material[1-10] can be attributed to the spontaneous generation of free electrons and holes following photo-absorption, thereby also resolving the apparent contradiction between initial reports of sizeable exciton binding energy values (30-50 meV) and recent reports of free carrier behavior [16,17,20,25,30-32] . The reason for the difference between our observations and previous estimates of the binding energy is primarily the combination of higher quality, much more crystalline films, and the use of very high magnetic fields. This has enabled us to measure multiple excitonic transitions which allow precise spectroscopic measurements, in contrast to previous works which have relied exclusively on measuring the 1s exciton and require assumptions to be made about the dielectric constant and effective masses[11-13, 17]. The effective mass values of ~0.1 $m_e$, which we have determined, are in good agreement with recent calculations [14,18,19,24] but are also significantly lower than the earlier experimental estimates [12,13].

[15]

## Methods

**Sample preparation**. The glass substrates were cleaned sequentially in hallmanex, acetone, isopropanol and $O_2$ plasma. The polycrystalline $CH_3NH_3PbI_3$ perovskite films were deposited in a nitrogen-filled glovebox following the interdiffusion preparation methods described previously[5]. In brief, a $PbI_2$ layer was first deposited on cleaned glass by spin-coating (speed/ramp = 6000rpm/6000 rpm/s, time = 35s) from a precursor solution of $PbI_2$ in DMF with concentration of 450 mg/ml, followed by drying at 70°C for 5 min. Then the MAI layer was deposited on dried $PbI_2$ film by spin-coating (speed/ramp = 6000rpm/6000 rpm/s, time = 35s) from a precursor solution of MAI in isopropanol with concentration of 50 mg/ml, followed by annealing at 100°C for 1 hour. The perovskite films were sealed by spin-coating a layer of the insulating polymer poly(methyl methacrylate) (PMMA; 10 mg/ml, speed/ramp = 1000rpm/1000 rpm/s, time = 60s) on top in order to ensure air-and moisture-insensitivity.

**Magneto-optical measurements** The measurements have been performed using 70 T long-duration and 150T short duration pulsed magnets in the high magnetic field laboratory in Toulouse. For the long pulsed measurements (~500 ms duration) the sample was immersed in liquid or gaseous helium in a cryostat. A tungsten halogen lamp was used to provide broad spectrum in the visible and near infra-red range. The absorption was measured in the Faraday configuration in which k, the wave propagation vector is parallel to the magnetic field B. A nitrogen cooled CCD array coupled to a spectrometer collected the light transmitted through the sample. The exposure time was 3 ms in order to limit variations in the magnetic field during acquisition. Thirty spectra were taken during a 70 T shot of the magnetic field. The magnetic field was measured using a calibrated pick-up coil. All spectra were normalized to both the incident intensity and by the zero field transmission to produce absolute and differential transmission spectra. For the short duration (10μs) pulsed measurements a series of diode and Ti-Sapphire laser lines was sent through and collected from the sample using fiber optics and detected using a fast (100MHz) silicon detector and high speed digital oscilloscope[33]. The sample was mounted inside a non-conducting helium flow cryostat and was cooled separately for each measurement. The magnetic fields were generated by a semi-destructive single turn coil system using 10mm coils[33,34].

[16]

**Acknowledgments**

The authors thank: Meso-superstructured Hybrid Solar Cells –MESO NMP-2013-SMALL7-604032 project. HJS thanks for funding the Engineering and Physical Sciences Research Council (EPSRC), the European Research Council (ERC-StG 2011 HYPER Project no. 279881). Dr. S. Stranks thanks Worcester College, Oxford, for additional financial support. This work was supported by EuroMagNETII under the EU contract No. 228043.

**Author contribution**

Atsuhiko Miyata, Anatolie Mitioglu, P.P, O.P and R.J.N collected and analysed the data. J. T-W. W. and S.D.S. prepared the samples. All authors contributed to the interpretation and the manuscript preparation. R.J.N. supervised and initiated the project.